\documentstyle[12pt]{article}

\large
\newcommand{\be}{\begin{eqnarray}}
\newcommand{\ee}{\end{eqnarray}}
\newcommand\del{\partial}

\newcommand\noi {\noindent}

\begin{document}
\setlength{\baselineskip}{21pt}
\pagestyle{empty}  
\vfill                                                      
\eject
\begin{flushright}
SUNY-NTG-95-53
\end{flushright}
\vskip 2.0cm 
\centerline{\large {\bf Two-Dimensional Solitons at Finite Temperature }}
\vskip 2.0 cm

\centerline{M. Kacir and I. Zahed} 
\vskip 1cm
\vskip .5cm
\centerline{Department of Physics}
\centerline{State University of New York at Stony Brook}
\centerline {Stony Brook, New York 11794-3800}
\vskip 2cm                                                                   
\centerline{\bf Abstract}
The partition function of two-dimensional solitons in a heat bath of mesons is 
worked out to one-loop. For temperatures large compared to the meson mass, the 
free energy is dominated by the meson-soliton bound states and the zero modes,
a consequence of Levinson's theorem. Using the Bethe-Uhlenbeck formula we 
compare the soliton energy-shift to the shift expected in the pole 
mass using a virial expansion. We construct the partition function
 associated to a fast moving soliton at finite temperature, and
 found that the soliton thermal inertial mass is no longer constrained
 by Poincare's symmetry. At finite temperature, the concept of
 quasiparticles is process dependent.

\vfill
\noindent
\begin{flushleft}
SUNY-NTG-95-53\\
December 1995
\end{flushleft}
\eject
\pagestyle{plain}

${\bf 1.}$
The statistical mechanics of mesons in the presence of solitons has been
studied extensively in the past using two-dimensional models \cite{TWO}. 
These systems arise naturally in physical settings involving polymeric 
structures \cite{POLY}, and have been used successfully to describe a 
variety of thermodynamical quantities \cite{THERMO}. 

At zero temperature, 
solitons provide an ideal set up for describing extended particles in a way 
that is fully consistent with the Poincare symmetry.
Meson-soliton and soliton-soliton scattering can be systematically
analyzed using a $1/\hbar$-expansion. In four dimensions, the semi-classical 
expansion is motivated by QCD in the large number of colors, and provide an 
interesting way of organizing the strong coupling expansion. Given the 
interest in finite temperature problems in QCD, it is natural to ask how 
meson-soliton description fetch at finite temperature. 

In this letter, we will address the finite-temperature issue in 
two-dimensional models, as a way to develope some insights
 to the behaviour of classical and extended objects in a heat
 bath of mesons, using semi-classical techniques. We start by
 constructing the partition function of a soliton in a heat bath of 
mesons. The free energy is calculated to one-loop, using the exact 
meson-soliton scattering amplitude. The large T behaviour of the 
free energy is generic, and solely determined by the number of
 meson-soliton bound states and zero modes, a direct consequence
 of Levinson's theorem. The relation to the soliton mass-shift
 at finite temperature is elucidated using the Bethe-Uhlenbeck
formula. Finally, we find that at finite temperature the 
energy-mass relation is no longer supported by the extended nature
 of the soliton. The relevance of these results to four-dimensional
 soliton models is discussed in our concluding remarks.

\vskip .5cm
${\bf 2.}$ Consider a field theory in two-dimensions with a given 
Lagrangian density
\be
{\cal L}=\frac{1}{2}\del _{\mu}\Phi 
\del ^{\mu }\Phi -U(\Phi)
\label{1}
\ee
For a class of time-translational potentials \cite{SINE}, the action 
associated to (\ref{1}) supports classical soliton or kink configurations 
with finite energy. They will be referred to as by $\Phi _{0}(x-X)$,
 where $X$ indicates that the finite energy configuration is
 space-translational invariant. The quantum theory associated to
 (\ref{1}) supports particle and meson states. The classical soliton
 configurations are precursors to the quantum particle states.
 Their characterization and properties will be assumed here \cite{REBBI}.

In the one-soliton sector, the meson-soliton dynamics can be organized 
systematically in $1/\hbar$. Indeed, by elevating $X$ to the rank of a
quantum variable, we have
\be
\Phi (x,t)=\Phi _{0}(x-X(t))+\phi(x-X(t),t)
\label{2}
\ee
the meson fluctuation $\phi $ is made transverse to the zero mode 
$\Phi _{0}^{\prime}$, to insure decoupling. The meson-nucleon Hamiltonian 
to order $\hbar^0$ reads
\be
H=M_{0}+\frac{1}{2}\int \!\left(\pi _{\phi}^{2}
+\phi ^{\prime 2}+U^{\prime \prime}(\Phi_{0})\phi ^{2}\right)
\label{3}
\ee
where $M_0=\int \! \Phi _{0}^{\prime}$ is the classical soliton mass.
The soliton kinetic energy $P^2/2M_0$ is of order $\hbar^{-1}$.

In the one-soliton sector the partition function associated to (\ref{3})
at temperature $T=1/\beta$, is given by 
\be
Z_{s}=\sqrt{M_{0}}{\rm e}^{-\beta M_{0}}Z_{\phi}
\label{4}
\ee
where, 
\be
Z_{\phi}=\left(\prod_{bs \neq zm}\frac{{\rm e}^{-\beta \omega _{n}/2}}
{1-{\rm e}^{-\beta \omega _{n}}}\right) 
\left(\prod_{sc }\frac{{\rm e}^{-\beta \omega _{n}/2}}
{1-{\rm e}^{-\beta \omega _{n}}}\right) 
\label{5}
\ee
The first product runs over the possible meson-soliton bound states
which are distinct from the zero modes (zm). The second 
product runs over the meson scattering states (sc). The energy modes 
$\omega_n$ diagonalize the $\hbar^0$ part of the Hamiltonian
(\ref{3}). Since the soliton kinetic energy has been ignored, 
(\ref{5}) is valid for temperatures $T<<M_0$. 
To account for vacuum renormalization, we will consider the
 ratio of $Z_s$ (\ref{4}) to $Z_0$, the partition function in
 the absence of the soliton. In the dilute gas 
approximation, the ratio $\hat{Z}_s=Z_s/Z_0$ provides the rationale
 for a virial expansion. This ratio will be used below.

To be able to count the number of meson states, we consider the system
in a spatial box of dimension $L$, with periodic boundary conditions.
 As a result, the meson-soliton phase-shifts $\delta_n (k)$ for a
 meson of energy $\omega_n=\sqrt{m_{\pi}^2 + k_n^2}$, satisfy
\be
Lk_{n}+\delta(k_{n})=2n\pi
\label{6}
\ee
As a result, we the pressure is just given by
\be
-\log \hat{Z}_{s} = &&+\beta\left(M_{0}-
\frac{\Lambda\delta(\Lambda)}{2\pi}
+\int\!\frac{dk}{2\pi}\delta^{\prime}(k)\frac{\omega}{2}\right) 
\nonumber\\&&+\sum _{bs \neq zm} 
\left(\frac{1}{2}\beta \omega _{n}+\log (1-{\rm e}^
{-\beta \omega _{n}})\right)
\nonumber \\ 
&&+\int _{\Lambda _{-}}^{\Lambda _{+}}\!\frac {dk}{2\pi} 
\delta ^{\prime}(k)
\log (1-{\rm e}^{-\beta \omega _{k}})
\label{7}
\ee
where $\Lambda_{\pm}=\pm (\Lambda -{\delta (\Lambda)}/{L})$. The 
ultraviolet cut-off for the soliton in free space is chosen to be 
$\Lambda=2N\pi/L$, where N is some fixed but large number. The ensuing 
logarithmic divergence in the mode sum, results in a finite 
renormalization of the soliton mass \cite{RENORM}.
If we denote it by $c$ ($c$ is zero for the sine-Gordon model) we have
\be
-\log \hat{Z}_{s}=&&+\beta \left(M_{0}-
\frac{\Lambda\delta(\Lambda)}{2\pi}
+c\right)\nonumber\\&&+\sum _{bs \neq zm}\left(\frac{1}{2}\beta 
\omega _{n}+\log
(1-{\rm e}^{-\beta \omega _{n}})\right)  
\nonumber \\ 
& &+\int _{\Lambda _{-}}^{\Lambda _{+}}\!\frac {dk}{2\pi} 
\delta ^{\prime}(k)
\log (1-{\rm e}^{-\beta \omega _{k}})
\label{8}
\ee
If we denote by $-\log Z_{T}$, the contribution of the meson
 fluctuations to the pressure in the one-soliton sector, then as 
\be
-\log Z_{T}=+\sum _{bs \neq zm}\log (1-{\rm e}^{-\beta \omega _{n}})
+\int _{- \Lambda } ^{\Lambda }\!\frac {dk}{2\pi} \delta ^{\prime}(k)
\log (1-{\rm e}^{-\beta \omega _{k}})
\label{9}
\ee
The hight temperature limit of (\ref{9}), $m \ll T < M_0$, is quoted
 in Table 1,
using the explicit form of the phase shifts for the Sine-Gordon model 
\cite{SINE} and the $\Phi^4$ model. These results can be understood 
independently of the detailed knowledge of the phase-shifts as
 we now show.

\vskip .5cm
${\bf 3.}$ To unravel the behaviour of (\ref{9}) in the high
 temperature regime $T \gg m_{\pi}$, we can expand (\ref{9})
 in $\beta=1/T$. A Taylor expansion of (\ref{9}) gives
\be
\log Z_{T}=-\sum _{bs \neq zm}\log \beta \omega _{n}
+\int _{-\Lambda }^{\Lambda }\!\frac {dk}{2\pi} 
\frac{k \delta (k)}{\omega ^{2}}
+n\log \beta m_{\pi } + {\cal O} (\beta m_{\pi} )
\label{10}
\ee
The integral over the phase shift in (\ref{10}) can be unwound using 
Jost functions \cite{NEWTON}. The result for the integral part is
\be
-\log\left(-im_{\pi}\dot{f}(-im_{\pi})\prod _{bs \neq zm}
\frac{m_{\pi}}{m_{\pi}-k_{n}}\right)
\label{11}
\ee
where the derivative of the Jost function 
$f(k) =S(k) = e^{2i\delta (k)}$ 
is related to the behaviour of the scattering amplitude in the 
lower half of the complex k-plane, at $k=-im_{\pi}$ and 
$k=-ik_{n}$ corresponding to
the locations of the bound states (bs) with the zero modes (zm)
 included. For instance,
\be
\dot{f}(-im_{\pi})=2(-im_{\pi})c_{+}c_{-}M_{0}
\label{12}
\ee
where $c_{\pm}$ are two constants to be determined (see below).

To see how (\ref{12}) comes about, a small digression into the 
scattering problem is necessary. For the normal modes, the
 scattering equation associated to (\ref{3}) reads
\be
\left[-\frac{\del ^{2}}{\del x^{2}}+U^{\prime \prime}(\Phi
_{0})-m_{\pi}^{2}\right]\ y_{\pm}(k,x)\
 = \ k^{2}\ y_{\pm}(k,x)
\label{13}
\ee
Here $y_{+}(k,x)$ and $y_{-}(k,x)$ are the two independent solutions
of the scattering equation with boundary conditions
$y_{+}(k,x\rightarrow +\infty)={\rm e}^{-ikx}$ and 
$y_{-}(k,x\rightarrow -\infty)={i}{\rm e}^{ikx}/2k$.
As in the three dimensional scattering radial equation the 
Jost function $f(k)$  relates these solutions through
\be
y_{-}(k,x)=\frac{i}{2k} y_{+}(-k,x)f(k)
\label{14}
\ee
Proceeding with analytic continuation in the lower half of the complex
plane of $k$ the discrete solutions (bs) of (13) occur for the zeros
of the Jost function. In particular $\Phi ^{\prime}_{0}(x)$ 
being the zero mode solution at $k=-im_{\pi}$, 
we have
\be
y_{\pm}(-im_{\pi},x)=c_{\pm} \Phi _{0}^{\prime}(x)
\label{15}
\ee

In terms of (\ref{11}-\ref{12}0 and (\ref{15}), the meson induced 
pressure in the one-soliton sector at high temperature (\ref{10}) 
reads
\be
-\log Z_{T} = +\log \left( -i \dot{f} (-im_{\pi}) T  \prod _{bs \neq
zm} \frac{\omega _{n}}{m_{\pi}-k_{n}} \right) + {\cal O}\left (\frac 
{m_{\pi}}T\right)
\label{16}
\ee
This equation is remarkable. It shows that to leading order in the 
temperature, the soliton pressure to one-loop is totally driven
 by the lowest lying states
(zero modes and bound states). We expect this result to extend to four 
dimensions \cite{KACIR}. The value of the Jost function $f(-im_{\pi})$
can be obtained by using (\ref{14}) along with the boundary 
conditions on $y_{\pm}$ to obtain $c_{\pm}$.
With $c_{\pm}$ we then make use of (11) to get $f(-im_{\pi})$. 
The only explicit form is contained in the zero mode $\Phi
_{0}^{\prime }(x)$. To gain more insight we work out the cases for 
both the Sine-Gordon and $\Phi ^{4}$ models. The results are displayed
 in Table 1. 
Both formula (9) and (15) are independently used to check the validity
of our manipulation. We point out that $-\log Z_{T}$ through (9) was 
worked out in \cite{MAKI}.

\vskip 1cm
\begin{center}
\begin{tabular}{|c|c|c|} \hline & & \\
{\rm Model}&{\rm sine-Gordon}&{\rm $\Phi ^{4}$} \\ \hline 
& &  \\
$U(\Phi)$&-$\frac{m^{2}}{g^{2}}(\cos g\Phi -1)$&$\frac{1}{2}
g^{2}(\Phi ^{2}-\frac{m^{2}}{g^{2}})^{2}$ \\ \hline 
& & \\
$\Phi _{0}(x)$&$\frac{4}{g}\arctan {\rm e}^{mx}$&$\frac{m}{g}\tanh
(mx)$ 
\\ \hline 
& & \\ 
$M_{0}$&$8m/g^{2}$&$4m^{3}/3g^{2}$ \\ \hline 
& & \\
b.s.&$-im_{\pi}=-im$& $-im_{\pi}=-2im;-ik_{1}=-im $ \\ \hline 
& & \\
$c_{+}$&$g/4m$&$g/4m^{2}$ \\ \hline 
& & \\ 
$c_{-}$&$-g/8m^{2}$&$-g/16m^{3}$ \\ \hline  
& & \\ 
$\dot{f}(-im_{\pi})$&$i/2m$&$i/12m$ \\ \hline  
& & \\ 
$-\log Z_{T}(15)$&$-\log2\beta m$&$-\log 4\sqrt{3}\beta m $ \\ \hline 
& & \\ 
$\delta (k)$& $2\arctan \frac{m}{k}$&$2(\arctan \frac{2m}{k}+\arctan 
\frac{m}{k})$ \\ \hline 
& & \\
$-\log Z_{T}(9)$&$-\log 2\beta m $&$-\log 4 \sqrt{3} \beta m $  \\ \hline 
\end{tabular}
\end{center}
\vskip 1cm
\centerline{Table 1.}

\vskip .5cm
${\bf 4.}$ Given the pressure $P_s=-{\rm ln} Z_T$, the energy of the 
one-soliton in the heat bath of mesons follows through
$E_s = -T\partial P_s/\partial T$. Using (\ref{9}), we have to one-loop
\be
E_{s}&=&\left( M_{0}-\frac{\Lambda \delta (\Lambda)}{2\pi}+c\right)
+\sum _{bs\neq zm}
\omega _{n}\left(\frac{1}{2}
+\frac{1}{{\rm e}^{\beta \omega _{n}}-1}\right)\nonumber \\ 
& &+\int _{-\Lambda}^{+\Lambda}\frac{dk}{2\pi}\delta ^{\prime}(k)
\frac{\omega _{k}}{{\rm e}^{\beta \omega_{k}}-1}
\label{17}
\ee
The first term in (\ref{17}) is the zero temperature mass-shift 
to order $\hbar^0$. In the Sine-Gordon model $c=0$, and the 
$-\Lambda\delta (\Lambda )/2\pi = -m/\pi$ \cite{DASHEN}. 
The temperature effects in (\ref{17}) are just the ones expected from 
meson thermal weights in phase space, which is a good check on (\ref{9}). 
From (\ref{17}) the soliton mass shift (or energy shift)
at finite temperature is just given by
\be
\delta M_{T}=\sum _{bs \neq zm}
\frac{\omega _{n}}{{\rm e}^{\beta \omega _{n}}-1}
+\int _{-\infty}^{+\infty}\! \frac{dk}{2\pi}\frac{\omega 
\delta ^{\prime}(k)}{
{\rm e}^{\beta \omega}-1}
\label{18}
\ee
For temperatures $T\gg m_{\pi}$ knowledge of the meson-soliton 
phase-shift is not needed. Indeed,
\be
\delta M_{T}=T(n-1)+T\frac{\delta(+\infty)-\delta(0^{+})}{\pi}
+{\cal O}(m\log \beta m)
\label{19}
\ee
Using Levinson's theorem \cite{NEWTON}, we obtain
\be
\delta M_{T}=-T+{\cal O}(m\log \beta m)
\label{20}
\ee
This result is generic to soliton models in two-dimensions. In the
temperature range $m_{\pi} \ll T \ll M_0$, the soliton mass shift as 
given by the energy definition, is negative and linear in T. 
The slope is controlled by (minus) the total number of zero modes.
 In (\ref{20}) the slope is just $-1$, since there is only one
 translational zero-mode in two-dimensions, in models without isospin.
The result (\ref{20}) can be directly checked from (\ref{18}) using
 the explicit forms of the phase shifts. The results are summarized
 in Table 2. Is the concept of a thermal 
soliton mass shift discussed in this section unique ?

\vskip .5cm
${\bf 5.}$ For an extended particle the mass definition is unique 
both at the classical and quantum level. The definition is process
 independent. 
It is the same whether we compute the energy of a soliton,
 its recoil mass or measure its time-like correlation functions.
 The uniqueness is insured by the fact that the 
Poincare symmetry is properly enforced to all orders in $\hbar$. 
At finite temperature, this is no longer the case as we now show. 

The soliton pressure and hence energy has provided us with one
 possible definition of the thermal soliton mass shift.
 This definition gives $\Delta 
M_T$ that is real to all orders in $1/\hbar$. Alternatively and 
in a dilute meson gas, a propagating soliton may receive a correction
 to its mass through local re-scattering processes. The latter cause
 the soliton pole mass to shift. The shift can be organized in
 powers of the pion density (virial expansion).
 To first order in the pion density,
\be
\delta M^*_{T}=-\int \!\frac{dk}{2\pi}\frac{t^{lab}}{2\omega}
\frac{1}{{\rm e}^{\beta \omega}-1}
\label{21}
\ee
where $t_{lab}$ is the forward scattering meson-soliton amplitude.
To get a simple estimate on (\ref{21}) at high temperature, 
consider a simple meson-fermion Yukawa-coupling ${\cal L} = 
f \overline{\psi} i\gamma_5 \gamma_{\mu} \psi\, \partial^{\mu} \pi$,
 where $\psi$ refers to a fermion of mass $M$. The result is 
$\delta M^*_T =f^{2}{m_{\pi}} T/{8M}$. The soliton shift at high 
temperature is subleading in $1/\hbar$ and linear in $T$. 
Since $M\sim /\hbar$, the mass shift is actually zero to order 
$\hbar^0$.

To understand the discrepancies between the mass-shift $\delta M_T$
(\ref{18}) provided by 
the energy-definition, and the mass-shift $\delta M_T^*$ 
(\ref{21}) provided by the 
pole-definition, we call upon the Bethe-Uhlenbeck formula
\be
\delta ^{\prime}(k)=\frac{d}{dk}\left(\frac{Re\
t(k)}{2k}\right)+\frac{i}{8k^{2}}(tt^{\prime \ast}
-t^{\prime}t^{\ast})
\label{23}
\ee
with the convention
\be
S={\rm e}^{i\delta (k)}=1-\frac{t(k)}{2ik}
\label{24}
\ee
In the absence of low-lying resonances, the quadratic term in the 
scattering amplitude can be ignored \cite{LANDAU}.
 In the present discussion, this is not possible since in the soliton
 sector there are low-lying zero modes, besides the 
zero modes (recoiling fermion in the intermediate state). Inserting 
(\ref{23}) into (\ref{17}) yields
\be
\delta M_T = &&+\sum_{bs \neq zm}
\frac{\omega _{n}}{{\rm e}^{\beta \omega _{n}}-1}
-\int \!\frac{dk}{2\pi}\frac{Re\ t}{2\omega}
\frac{1}{{\rm e}^{\beta \omega}-1}
\nonumber \\
&&+\int \!\frac{dk}{2\pi}\frac{Re\ t}{2}\frac{\beta 
{\rm e}^{\beta \omega}}{({\rm e}^{\beta \omega}-1)^{2}}
\nonumber \\
&&+\int \!\frac{dk}{2\pi}\frac{i}{8k^{2}}
(tt^{\prime \ast}-t^{\prime}t^{\ast})\frac{\omega}
{{\rm e}^{\beta \omega}-1}
\label{25}
\ee
The first term ($\delta M_{\rm bs}$)
is the contribution of the bound states viewed 
as sharp resonances, the second term ($\delta M_{\rm sol}$)
is the (real part) analog of (\ref{21}), the third term 
($\delta M_{entro.}$)
is the entropy-counterpart of the second term, and the fourth term 
($\delta M_{quadra.}$) contains the contributions from 
possible resonances.

In the high temperature limit $\delta M_{sol.}=-T/2$ for both 
the Sine-Gordon and the $\Phi^4$ models. 
The contributions $\delta M_{entro.}+\delta
M_{quadra.}$ are respectively $-T/2$ and $-3T/2$ for the sine-Gordon
and the $\Phi ^{4}$ models. In Fig. 1 we plot these various
contributions to the soliton mass shift in the sine-Gordon model for
the low and high temperature regimes. 
Clearly the quadratic term following from the Bethe-Uhlenbeck formula
cannot be ignored when assessing the mass shift using the energy 
definition. 

The discrepancy between $\delta M_{sol.} = -T/2$ at high temperature 
following from (\ref{25}) and $\delta M_T^* \sim T$
 following from (\ref{21}) using a 
Yukawa-coupling can be traced back to the schematic nature of the 
scattering amplitude.
 Indeed, in weak coupling limit, the meson-soliton $t$-matrix is 
known exactly. It contains a Born term with intermediate bound and 
continuum states plus a seagull term $K$, following from an equal-time 
commutator \cite{LIANG}. The contribution of the bound 
states (including zero modes) to the mass shift (\ref{25})
\be
\delta M^{0}_{sol.}=-\int \!
\frac{dk}{2\pi} \frac{t_{0}^{lab}}{2\omega }
\frac{1}{{\rm e}^{\beta \omega}-1}
\label{26}
\ee
where
\be
t_{0}^{lab}=\frac{-k^{4}}{M\omega ^{2}}f_{0}^{2}(k^{2})
\label{27}
\ee
Here $f_{0}(k^{2})$ is the meson-soliton Yukawa form factor 
\cite{LIANG}. 
We obtain, for both the Sine-Gordon and $\Phi ^{4}$, $\delta
M_{sol}^{0}=+T/2$. Obviously some important contributions still need
to be evaluated. For instance $\delta M_{K}$ and $\delta M_{cont.}$
respectively associated to the $K$ (seagull) and $t_{cont.}$
(continuum+excited states) scattering amplitudes \cite{LIANG}. 
The results are displayed in Table 2.
 In Fig. 2, we show the respective contributions for 
finite $T$ in the Sine-Gordon model. While the soliton-meson 
Yukawa interaction 
drives the mass contribution (\ref{26}) up, the seagull part 
drives it down.
We note that in both the Sine-Gordon model and the $\Phi^4$ model, 
\be
\delta M_{sol.} = \delta M_{sol}^0 + \delta M_K + 
\delta M_{cont.} = -T +{\cal O} \left(\frac T{m_{\pi}} \right)
\label{28}
\ee
in agreement with (\ref{20}), implying that the 
bound-state, entropy, and quadratic contributions as illustrated
 in (\ref{25}) cancel out at high temperature. In general, however,
 the two-definitions of the soliton mass shift (\ref{17}) and
 (\ref{21}) (or equivalently (\ref{26})) are not necessarily the same.

\begin{center}
\begin{tabular}{|c|c|c|} \hline & & \\
{\rm Model}&{\rm sine-Gordon}&{\rm $\Phi ^{4}$} \\ \hline 
& &  \\
$\delta M_{T}(17)$&-$T/2$&$-T/2$ \\ \hline 
& & \\
& & \\
$f_{0}(k^2)$&$
\frac{2\pi}{gk^{2}}\frac{\omega ^{2}}{\cosh \pi
k/2m}$&$-\frac{\pi}{gk}\frac{\omega ^{2}}
{\sinh \pi k/2m}$
 \\ \hline 
& & \\
& & \\
$\delta M_{T}^{0}(25)$&$+T/2$&$+T/2$ \\ \hline 
& & \\
K&$4m$&$12m$ \\ \hline
& & \\
$\delta M_{K}$&$-T$&$-3T/2$ \\ \hline 
& & \\ 
$\delta M_{cont.}$&$0$&$+T/2$ \\ \hline  
& & \\ 
$\delta M_{sol.}$&$-T/2$&$-T/2$ \\ \hline  
\end{tabular}
\end{center}
\vskip 1cm
\centerline{Table 2.}

\vskip .5cm
${\bf 6.}$ The above discrepancy may imply that in a heat bath, 
the dispersion relation of a soliton no longer follows
 the general lore of the vacuum, where 
Poincare's invariance implies
\be
E_s (v) = M_s  + \frac 12 M_s v^2 + {\cal O} (v^2)
\label{28}
\ee
with $M_s = M_0 + \Lambda\delta (\Lambda )/2\pi + c$, is the one-loop 
corrected soliton mass \cite{LEE,JAIN}. To see how the soliton 
disperses at finite temperature, consider a fast moving soliton
 in a periodic box of size $L$.
 A rerun of the above arguments, show that the analog of (\ref{6}) is
\be
L\gamma (k_{n}+\omega _{n}v)+\delta _{k_{n}}=2n \pi
\label{29}
\ee
where the energy $\omega_n$ and the momentum $k_n$ of a meson are in
 the soliton rest frame. Here $\gamma=1/\sqrt{1-v^2}$ is the Lorentz 
contraction factor. The partition function for a fast moving soliton 
follows from a proper quantization of the meson-soliton system with
 boosted configurations \cite{JAIN}. To one-loop, the result is
 (the $bs \neq zm$ contribution is dropped for simplicity)
\be
-\log Z_{s}(v)=+\beta \gamma M_{0}+\sum_{n}\left(\frac{1}{2}\beta
\gamma \omega _{n}+\log(1-{\rm e}^{-\beta \gamma 
\omega _{n}})\right)
\label{30}
\ee
The partition function of a boosted soliton follows from the one
 at rest (\ref{8}) by changing $M_0\rightarrow \gamma M_0$ and 
$\omega_n\rightarrow \gamma\omega_n$.
 The scattering mesons around a fast moving extended soliton 
are Doppler-shifted.  Subtracting the vacuum part for the moving
 system and after mass renormalization,  we obtain the reduced
 partition function
\be
-\log \hat{Z}_{s}(v)=+\beta \gamma (M_{0}-
\frac{\Lambda \delta (\Lambda
)}{2\pi}+c)
+\int _{\Lambda_{l}}^{\Lambda _{u}}\!
\frac{dk}{2\pi}\delta ^{\prime
}(k)\log(1-{\rm e}^{-\beta \gamma \omega _{k}})
\label{31}
\ee
The cutoffs $\Lambda _{u,l}$ are the same as in the zero-temperature 
case \cite{JAIN}. The result (\ref{31}) differ from the one discussed
 by Maki and Takayama \cite{MAKI} by a factor of $-\beta \gamma k v$
 absent in the exponential of (\ref{31}).
 The discrepancy lies in the enforcement of the boundary conditions 
on the scattering mesons in the presence of a fast moving soliton.
 Following \cite{JAIN}, our boundary conditions (\ref{29}) reflect on 
the fact that the soliton moves at a speed $v$ in a periodic box of 
length $L$.
From (\ref{31}), it follows that the moving soliton obeys the following 
dispersion relation
\be
E_{s}(v)= \gamma \left(M_{0}-\frac{\Lambda \delta (\Lambda )}{2\pi}
+c\right)-\int _{\Lambda_{l}}
^{\Lambda _{u}}\!\frac{dk}{2\pi} \delta ^{\prime }(k)
\frac{\gamma\omega_{k}}{{\rm e}^{-\beta \gamma \omega _{k}}-1}
\label{32}
\ee
For small velocities
\be
 E_{s}(v)=E_{s}(T) + \frac{1}{2} E_K (T) v^{2} +{\cal O} (v^2)
\label{33}
\ee
The thermal inertial mass $E_K  (T)$ differs from $E_s (T)$ and provides
for another definition of the soliton mass-shift at finite temperature.
Its high temperature limit is different from the ones discussed above.

\vskip .4cm
${\bf 7.}$ We have analyzed the partition function of two-dimensional 
solitons to one-loop. The soliton mass-shift following from the 
energy-definition was found to asymptote $-T$ independently of
 the model used.  This result is generic, and only conditioned
 by the number of zero modes. 
By construction, the energy-definition yields a mass-shift 
that is always real. We have also analyzed the soliton mass-shift 
following from the pole-mass definition. The result is generally
 complex, and reflects on the possibility of particle absorption
in the heat bath. Use of the Bethe-Uhlenbeck formula revealed that the 
energy-mass definition contains the real part of the pole 
mass-definition along with other contributions.
 In the high temperature limit, these contributions 
cancel out. Our considerations have shown the importance of the use
 of form factors, seagull as well as continuum in the scattering 
amplitude. A simple Born approximation yields erroneous results.
Finally, using boosted soliton configurations, we have shown that
the soliton dispersion relation at finite $T$ leads yet to a 
third mass-shift definition using the soliton kinetic energy in
 a heat bath.

Mass-shifts of extended particles at finite temperature provides 
for an important way to parameterize their properties at low
temperature. While the concepts are process dependent, they reflect
 on important aspects of the interactions of the quasiparticles in
 the thermal state. The energy-mass definition captures the
 essentials of the bulk  state relevant for transport properties.
 The pole-mass definition through its real and imaginary part 
provides for the essentials of the quasiparticle absorption and shifts.
Finally, the kinetic-mass definition provides for parts of the 
quasiparticle dispersion. In particular in the use of dispersion 
theories, it is important whether pole-masses or inertial-masses are 
used. The mass-shifts are found to be sensitive to be very sensitive 
to the various parts of the scattering amplitude 
(Born, seagull, continuum). Some of these findings may be important
 when analyzing thermal shifts of hadrons in four 
dimensional theories, whether using effective Lagrangians or QCD 
inspired models.

\vskip 1cm
{\bf \noindent  Acknowledgements } \\ 
\noi 
This work was supported in part by the 
US Department of Energy under Grant No. DE-FG-88ER40388.

\newpage
\vglue 1cm

\bibliographystyle{aip}
\bibliography{vacref}

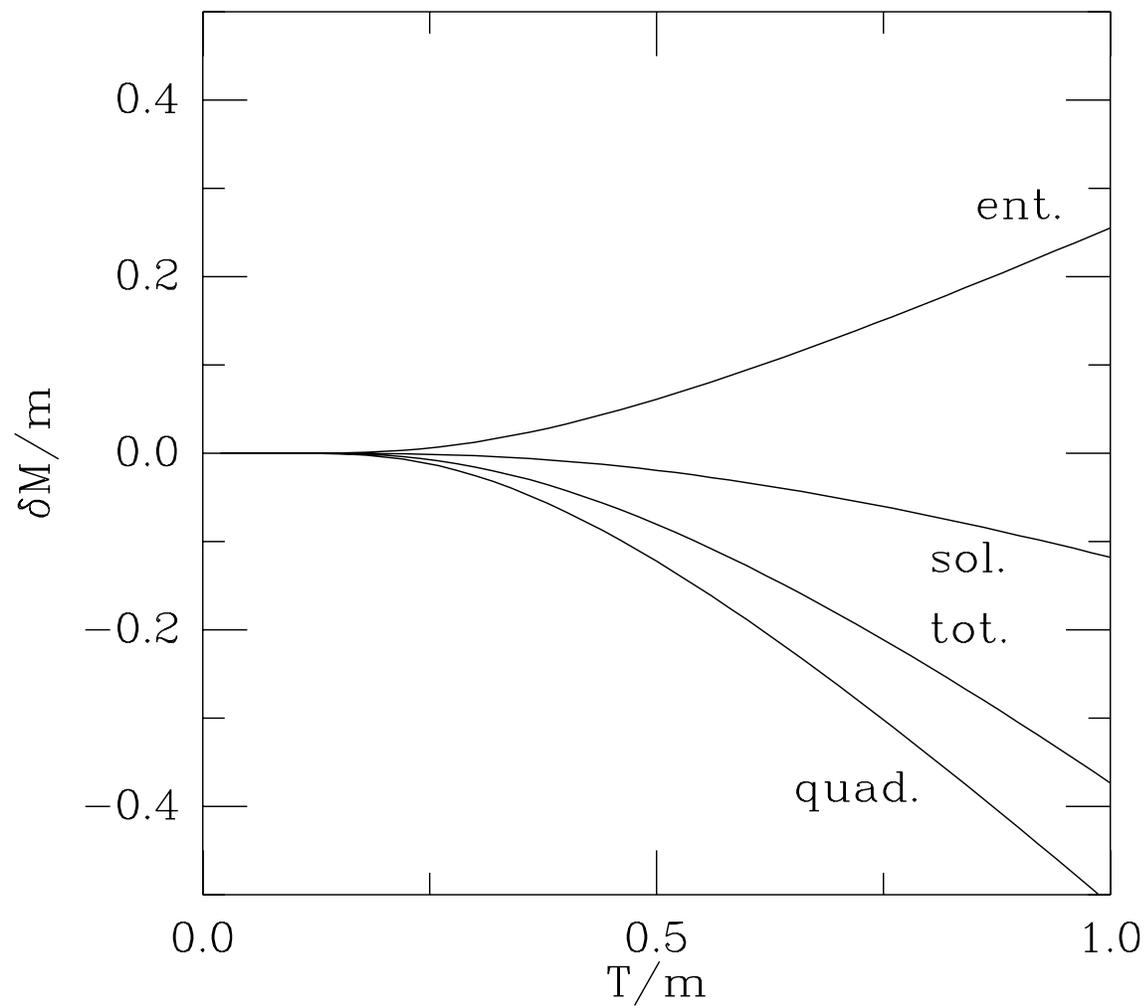
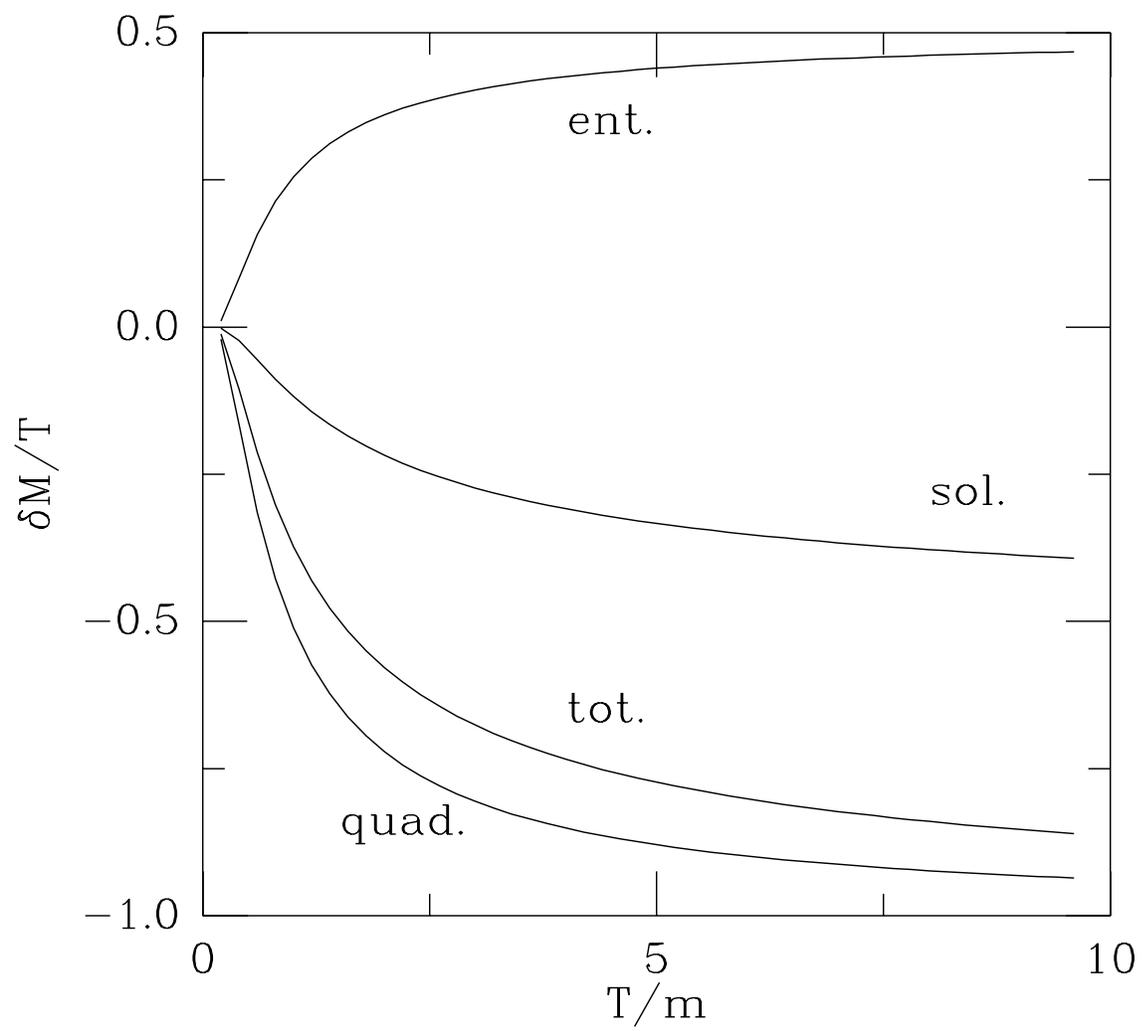
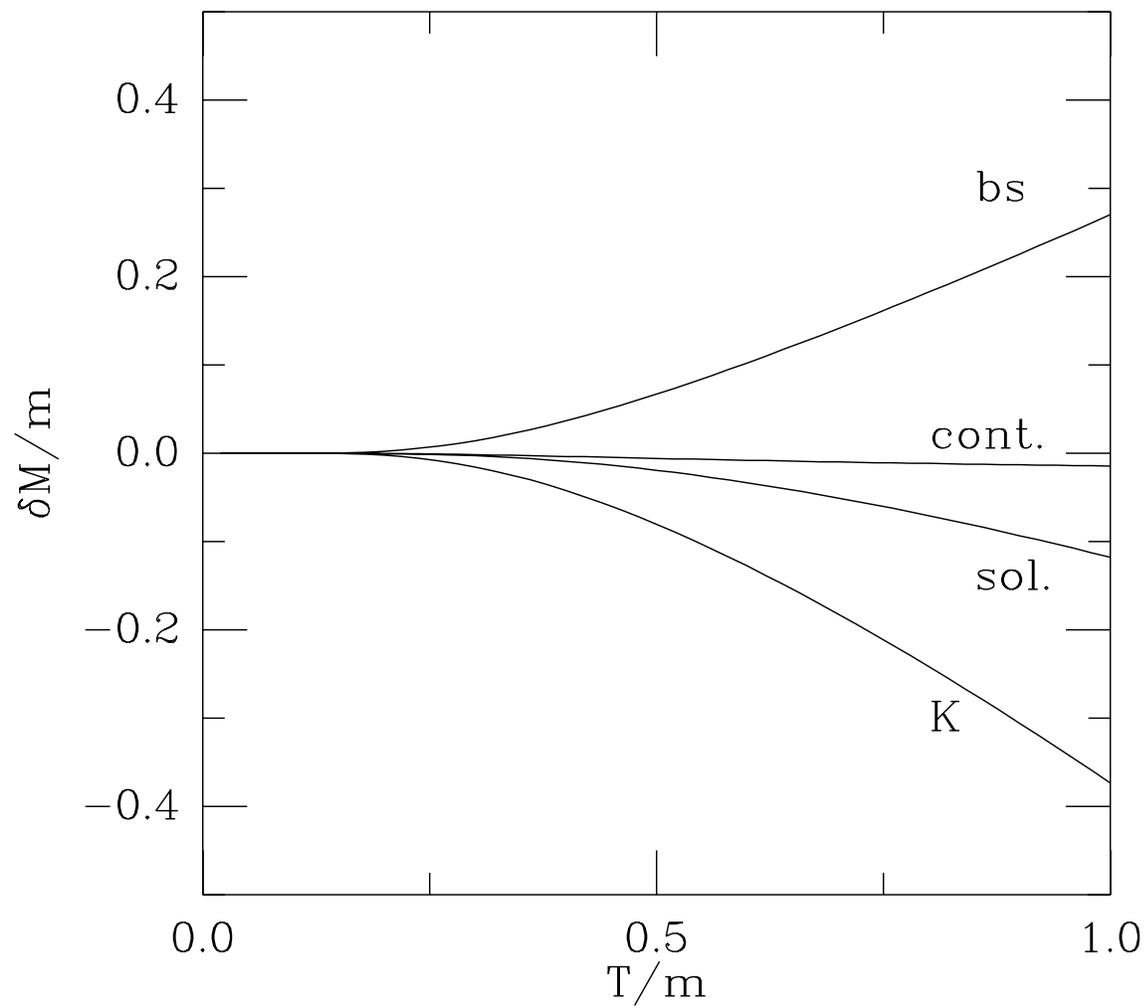
\end{document}